\documentstyle[preprint,aps,prb]{revtex}

\begin{document}

\draft

\title{ Diffusion Monte Carlo study of two-dimensional liquid
$^{\bf 4}$He}

\author{ S. Giorgini, J. Boronat and J. Casulleras }

\address{Departament de F\'{\i}sica i Enginyeria Nuclear, Campus Nord
B4--B5, \protect\\ Universitat Polit\`ecnica de Catalunya, E--08028
Barcelona, Spain}


\maketitle

\begin{abstract}

The ground-state properties of two-dimensional liquid $^4$He at zero
temperature are studied by means of a quadratic diffusion
Monte Carlo method. As interatomic potential we use a revised
version of the HFDHE2 Aziz potential which is expected to give a better
description of the interaction between helium atoms. The equation of
state is determined with great accuracy over a wide range of
densities in the liquid phase from the spinodal point up to the freezing
density. The spinodal decomposition density is estimated and other
properties of the liquid, such as radial distribution function, static form
factor, momentum distribution and density dependence of the condensate
fraction are all presented.

\end{abstract}


\narrowtext

\section{Introduction}

In recent years a great deal of interest has been devoted in the
literature to the study of quantum boson liquids in restricted
geometries.$^{1-15}$ From a theoretical point of view thin films of
liquid $^4$He adsorbed on different solid substrates have been
studied using variational techniques, based on the
hypernetted-chain/Euler-Lagrange (HNC/EL) theory,$^{1-3}$
Density Functional theories
(DFT),$^{4-6}$ and also Monte Carlo techniques.$^{12,13}$ According
to these investigations, liquid $^4$He films
display different behaviours depending on the strength of the
substrate potential.

In the case of weak-binding and long-range potentials, such as
some alkali metal substrates, the helium film does not exhibit at all
two-dimensional characteristics. By adding $^4$He atoms to the
substrate surface, three-dimensional clusters form and helium
uniformly covers the surface only for coverages many atoms thick
(prewetting transition). For some substrates a stable coverage is
found only for infinitely thick films (wetting transition).

More interesting, from the point of view of the role of
dimensionality, is the case of substrates with deep and narrow
potential wells for which the degree of freedom perpendicular to the
surface is practically ``frozen out" and low-coverage films are
stabilized at the surface forming two-dimensional (2D) systems. To
this category belong graphite, solid H$_2$ on glass
and some alkali metal substrates like Li and Mg.  On these substrates
the growth of the first liquid layers is predicted to proceed via
layering transitions:$^{1,3,4}$ by increasing the surface coverage
single atomic monolayers develop and become stable
one on the top of the other. This prediction has been confirmed
experimentally for helium on graphite by heat capacity$^{8}$
and third sound measurements.$^{9}$

Variational calculations of ground-state and dynamic properties of
single
$^4$He monolayers at zero temperature reveal a striking 2D behaviour
over a
wide range of coverages.$^{1,2}$ Such calculations seem to indicate
that single monolayers of liquid helium adsorbed on strong-binding
substrates represent physical realizations of 2D homogeneous
quantum liquids.

A condition which can be relevant for theories aiming to
describe $^4$He films is that they possess the correct limiting
behaviour of  homogeneous three-dimensional and two-dimensional
systems. Homogeneous three dimensional
(3D) $^4$He is a well studied system both experimentally and
theoretically.  In the case of homogeneous 2D
$^4$He no experimental results are available and ``exact"
results coming from {\it ab initio} calculations can be extremely
useful.

{}From the point of view of Monte Carlo simulations, $^4$He in confined
geometries has already been the object of several studies.$^{10-15}$
Variational Monte Carlo techniques have been applied to the study of
both inhomogeneous films on substrates$^{12}$ and more extensively to
homogeneous 2D $^4$He, where the simulation is easier.$^{10,15}$
However, these
methods rely heavily on the choice of the variational wavefunction
used in the simulation.

{}From the side of more exact Monte Carlo methods, thin $^4$He
films on molecular hydrogen surfaces at low temperature have been
recently studied by Wagner and
Ceperley$^{13}$ using Path Integral Monte Carlo (PIMC) techniques.
Strictly two-dimensional $^{4}$He has also been the object of Monte
Carlo calculations.$^{11,14}$ At
$T=0$ Whitlock {\it et al.}$^{11}$ have calculated, using Green's
Function Monte Carlo (GFMC),
the equation of state and other ground-state properties of liquid and
solid $^4$He in two dimensions, giving an estimate of the freezing
and melting densities. At finite temperature, PIMC
techniques have been employed by Ceperley {\it et al.}$^{14}$ to
investigate the
superfluid transition, which in 2D belongs to the Kosterlitz-Thouless
universality class.

In the present work we present a Diffusion Monte Carlo (DMC)
calculation of homogeneous liquid $^4$He in two dimensions at zero
temperature. Our main purpose is to provide an exhaustive
calculation of the equation of state and other relevant properties of
the ground state such as distribution function, structure form
factor, momentum distribution and condensate fraction for
a wide range of densities within the liquid phase.
In particular, we have performed a detailed analysis of the equation of
state
in the region of negative pressures in order to give an accurate
estimation
of the spinodal density of the 2D $^4$He liquid. The spinodal
density is
defined as the density below which the system becomes macroscopically
unstable against density fluctuations.  Its determination is
relevant for the
physics of $^4$He monolayers on strong-binding substrates because the
spinodal density coincides with the coverage at which the uniform
monolayer breaks down into 2D clusters in equilibrium with the vacuum,
and it is thus the density at which the first layering transition
occurs.$^{1,3}$

In our DMC simulation we have used as interatomic potential a revised
version of the HFDHE2 Aziz potential$^{16}$ which was proposed
by Aziz {\it
et al.}$^{17}$ in 1987 and is known as HFD-B(HE) potential.
This renewed
interatomic potential was recently used by two of us$^{18}$
(J.B. and J.C.)
to study 3D bulk helium at zero temperature and it has been proven to
give results for the density dependence of the pressure and the
system compressibility which are closer to experimental data than the
older Aziz potential. We are thus inclined to think that it gives also
better results for two-dimensional $^4$He.

The structure of the paper is as follows: in Sec. II we outline
briefly the quadratic Diffusion Monte Carlo method used to solve the
Schr\"odinger equation. In Sec. III we present results on both the
equation of state in the liquid phase and on the
distribution and structure functions.  In Sec. IV are collected the
results for the density dependence of the condensate fraction, which
differ substantially from the previous analysis by Whitlock and
coworkers,$^{11}$ and of the momentum distribution for different
values of
the density. A brief discussion and conclusions are included in Sec.
V.

\section{Computational method}

The aim of Diffusion Monte Carlo algorithms is to solve the
time dependent Schr\"odinger equation of a system of N particles in
imaginary time
\begin{equation}
-\frac{\partial\Psi({\bf R},t)}{\partial t} = (\hat{H} - E)
\Psi({\bf R},t) \ ,
\label{sch}
\end{equation}
where ${\bf R}=({\bf r}_1,...,{\bf r}_N)$ is the configuration vector
for the positions of the N particles and $t$ is measured in units of
$\hbar$.  In Eq. (\ref{sch}) the Hamiltonian $\hat{H}$ has the
usual form
\begin{equation}
\hat{H} = - \frac{\hbar^2}{2m} \nabla_{\bf R}^2  +  V({\bf R}) \ ,
\label{ham}
\end{equation}
and $E$ is a parameter representing an energy shift. Provided the
initial wavefunction $\Psi({\bf R},t=0)$ has a nonzero overlap with
the ground state $\Phi_0({\bf R})$, the solution of Eq. (\ref{sch})
gives exactly the ground state
wavefunction in the asymptotic limit of large times: $\Psi({\bf R},t
\rightarrow\infty) = \Phi_0({\bf R})$.

When numerically integrating Eq. (\ref{sch})
importance sampling techniques are employed to guide quickly the
solution
towards the ground state. To this aim the Schr\"odinger equation
(\ref{sch}) is rewritten for the function
\begin{equation}
f({\bf R},t) = \psi_T({\bf R}) \Psi({\bf R},t) \ ,
\label{fun}
\end{equation}
where $\psi_T({\bf R})$ is a time independent trial function. One
gets
\begin{equation}
-\frac{\partial f({\bf R},t)}{\partial t}= - D \, \mbox{\boldmath
$\nabla$}_ {{\bf R}}^2 f({\bf R},t) + D \, \mbox{\boldmath
$\nabla$}_{{\bf R}}( {\bf F}({\bf R}) f({\bf R},t) )+(E_L({\bf R})-E)
\,f({\bf R},t) \equiv \hat{A} \, f({\bf R},t) \ .
\label{dmc}
\end{equation}
Eq. (\ref{dmc}) has the form of a classical diffusion equation for
the distribution function $f({\bf R},t)$. The first term describes
the
diffusive process with diffusion constant $D=\hbar^2/2m$. The
second
term contains the drift force
\begin{equation}
F({\bf R}) = 2 \psi_T({\bf R})^{-1}\nabla_{\bf R} \psi_T({\bf R}) \ ,
\label{df}
\end{equation}
which drives the system towards the region in configuration space
where the trial function $\psi_T({\bf R})$ is relevant.  The third
term in Eq. (\ref{dmc}) represents a branching term which depends on
the local energy $E_L({\bf R})=\psi_T({\bf R})^{-1}\hat{H} \psi_T({\bf
R})$.

By introducing the time dependent Green's function
\begin{equation}
G({\bf R}',{\bf R},\Delta t) = \langle {\bf R}' \,| \,
e^{-\hat{A}\Delta t}
\, | \, {\bf R} \rangle
\label{gf}
\end{equation}
the solution of Eq. (\ref{dmc}) can be written formally as
\begin{equation}
f({\bf R}^{\prime},t+\Delta t)=\int d {\bf R} \ G({\bf
R}^{\prime},{\bf R}, \Delta t) f({\bf R},t) \ .
\label{green}
\end{equation}
If the Green's function $G({\bf R}',{\bf R},\Delta t)$ is known for
infinitesimal time steps $\Delta t$, the asymptotic solution for
large times $f({\bf R},t\rightarrow\infty)$ can be obtained by
solving iteratively Eq. (\ref{green}).

In the most usual implementations of DMC algorithms to solve
the many
body Schr\"odinger equation, $G({\bf R}',{\bf R},\Delta t)$
is approximated
up to order $\Delta t$ for small time steps.$^{19}$ In
this case, the obtained energy of the ground state depends
linearly on
the time step $\Delta t$ and several calculations for different
values of $\Delta t$ are needed in order to extrapolate the
correct
result in the limit $\Delta t\rightarrow 0$.

Recently a quadratic DMC algorithm$^{20}$ has been
proven to work efficiently in the description of $^4$He
droplets$^{21}$ and
bulk 3D liquid $^4$He.$^{18}$  This method relies on an
expansion of
the Green's function $G({\bf R}',{\bf R},\Delta t)$ up to
order $(\Delta
t)^2$, which generates a quadratic dependence of the energy
eigenvalue
on $\Delta t$, permitting thus to avoid the extrapolation to the
limit $\Delta t\rightarrow 0$ by working with a single time
step. More details concerning this procedure can be found
in Ref. 18.

In the present work we have used in all the calculations
$\Delta t =
2\times 10^{-3}\tau$, where $\tau$ is the appropriate time unit:
$\tau = m\sigma^2/2\hbar^2$ ($\sigma = 2.556$\AA). Time
step analysis
have been performed
at the density $\rho = 0.275 \sigma^{-2}$, close to the liquid
equilibrium density, and at $\rho = 0.400
\sigma^{-2}$, near the freezing density, and no changes
have been observed in the results for the energy eigenvalue by
reducing the time step $\Delta t$.

The mixed
estimator $\langle \psi_T \mid \hat{O} \mid \Phi_0 \rangle$
of a generic
operator $O$, is the direct output of DMC algorithms. If the
operator $\hat{O}$ commutes with the Hamiltonian of the system,
the mixed
estimator coincides with the pure expectation value on the ground
state $\langle \Phi_0
\mid \hat{O} \mid \Phi_0 \rangle$. If, on the contrary, the operator
$\hat{O}$ does not correspond to a conserved quantity, its
expectation value on the ground state can be obtained by means of
a linear extrapolation$^{22}$
\begin{equation}
\langle \Phi_0 \mid \hat{O} \mid \Phi_0 \rangle = 2 \langle \psi_T
\mid \hat{O} \mid \Phi_0 \rangle - \langle \psi_T \mid \hat{O} \mid
\psi_T \rangle \ .
\label{lex}
\end{equation}
The above method to get pure estimators is the most widely used in
DMC simulations which must thus be supplemented by a variational
Monte Carlo (VMC) calculation to determine the variational
expectation value $\langle \psi_T \mid \hat{O} \mid \psi_T
\rangle$.
The linear extrapolation (\ref{lex}), which is obtained by
writing the
ground state wavefunction $\Phi_0$ as $\Phi_0({\bf R}) =
\psi_T({\bf R}) + \delta \psi({\bf R})$, is correct to linear order
in the functional variation $\delta \psi$.

To go beyond this approximation, removing the dependence on the trial
wavefunction, several algorithms have been proposed to
calculate pure estimators.$^{23,24}$ On this line an
algorithm based on ``forward walking"
has been recently presented by two of us$^{25}$ (J.  C. and J. B.) which
is
easy to incorporate in the original Monte Carlo algorithm and allows the
calculation of pure expectation values of coordinate operators
$\hat{O}({\bf R})$ with
satisfactory stability and reliability.  The method applied to bulk
3D liquid helium$^{25}$ gives values for the particle-particle
distribution
function and static structure factor in very good agreement with
experimental results and without any significant dependence on
the function used as importance sampling. The results given in the
present work for the potential energy, distribution and structure
functions are calculated by using this method.

Another important parameter in the calculation is the population of
walkers $n_w$, which represents the number of points ${\bf R}_i$ in
configuration space at which the distribution function $f({\bf R},t)$
is sampled. In our calculations we have used a mean walker
population of $n_w = 450$. At the densities $\rho = 0.275
\sigma^{-2}$ and $\rho = 0.400
\sigma^{-2}$ the walker population was increased to $n_w = 900$
and no appreciable change in the results was observed.

Our simulation box contains 64 particles. At the 2D saturation
density $\rho_0 \simeq 0.284 \sigma^{-2}$ this corresponds to a
simulation box length of $\sim 38$ \AA, roughly 14 times larger than
the mean interparticle distance at this density. Such
box size is large enough to neglect finite volume effects. In fact, in
3D bulk liquid $^{4}$He finite size effects have been proven to be
negligible already for a box length of 8 times larger than the mean
interparticle distance.$^{18}$

Finally, as importance sampling we have used a simple McMillan$^{26}$
two-body trial function
\begin{equation}
\psi_T({\bf R}) = \prod_{i<j} \exp \, \left(
-\frac{1}{2} \left(\frac{b}{r_{ij}} \right)^5 \right) \ .
\label{tf}
\end{equation}
We have taken for all the densities $b = 1.205
\sigma$. This value, which minimizes the energy at the
equilibrium density, coincides with the one found by Whitlock {\it
et al.}$^{11}$ in their VMC calculation.
For the highest density
calculated, $\rho = 0.420 \sigma^{-2}$, we have also used as importance
sampling the two-body function proposed by Reatto$^{27}$
\begin{equation}
\psi_T({\bf R}) = \prod_{i<j} \exp \, \left(
-\frac{1}{2} \left( \frac{b}{r_{ij}} \right)^5 - \frac{L}{2} \exp
\left(- \left(\frac{r_{ij}-\lambda}{\Lambda}\right)^2\right)\right),
\label{tf1}
\end{equation}
with $L=0.2$, $\lambda=2.0\sigma$, $\Lambda=0.6\sigma$ and
$b=1.225\sigma$.
Although this trial function gives at the density
$\rho=0.420\sigma^{-2}$ a
VMC energy lower than the McMillan function (\ref{tf}), no
appreciable change in the results of the DMC simulation is found.

\section{Results}

In this section, we present our numerical results for the energy and
structure properties of the ground state. First we analyze
the equation
of state in the liquid phase, comparing our results with
the ones presented
in Ref. 11. Ground-state properties such as the radial
distribution function and the static structure factor are
also discussed
in the first subsection. The second subsection contains our results
concerning the density dependence of the condensate fraction and the
momentum distribution.

\subsection{Equation of state and structure properties}

In Ref. 11, the equation of state for both the liquid and the
solid phase of 2D homogeneous $^4$He has been investigated
by means of VMC
and
GFMC techniques. The interatomic potential employed in this
study is the
two-body HFDHE2 Aziz potential.$^{16}$ The estimated value for the
liquid
freezing density from the GFMC calculation is
$\rho_l = 0.443 \sigma^{-2}$.
For the sake of consistency we have calculated the energy
per particle using
the Aziz potential at two densities in the liquid phase:
at $\rho = 0.275
\sigma^{-2}$, close to the equilibrium density, and at $\rho = 0.400
\sigma^{-2}$, near freezing. The values for the energy
obtained from our DMC calculation are in good agreement with the GFMC
results reported in Ref. 11 for the same two densities.

Once the equivalence between our DMC algorithm and the GFMC algorithm
employed in Ref. 11 has been tested by using the same
interatomic potential, we have proceeded to the study of
the equation of
state in the liquid phase with the HFD-B(HE) potential proposed
by Aziz {\it et al.}$^{17}$
(henceforth referred to as Aziz II). In Table I we present the results
of our DMC calculations
for the total, potential and kinetic energy per particle for some of the
densities calculated. As discussed in Sec. II the values of the potential
energy per particle
have been obtained by employing the algorithm for pure
estimators.$^{25}$

The Aziz II potential is slightly more attractive than the Aziz
potential and the binding energies at the different densities are
therefore
somewhat larger with respect to the ones obtained in Ref. 11.
For example, at $\rho = 0.275 \sigma^{-2}$ near to equilibrium the
energies per particle
are $E/N = -0.8519 \pm 0.0044$ K and $E/N = -0.8950 \pm 0.0019$ K
for the
Aziz and Aziz II potentials respectively. In
Fig. 1 we show our DMC results for the equation of state in the
liquid phase
together with the GFMC results of Ref. 11.

The equation of state of 2D liquid $^4$He is usually fitted by using a
polynomial cubic function of the form
\begin{equation}
e = e_0 + B\left(\frac{\rho-\rho_0}{\rho_0}\right)^2 + C
\left(\frac{\rho-\rho_0}{\rho_0}\right)^3 \ ,
\label{fit}
\end{equation}
where $e=E/N$ and $\rho_0$ is the equilibrium density. In Table II we
report the values of the parameters which best fit our results and
we compare them with the values reported in Ref. 11.
The values for the equilibrium density are very close, whereas
appreciably different are the values for the $B$ and $C$ parameters.
These differences affect the predictions for the density
dependence of the surface pressure and compressibility, as well as
the estimation of the spinodal density. The two fits are shown
in Fig. 1
together with the Monte Carlo data. The cubic polynomial fit
(\ref{fit})
fits our data rather well and no significant improvement in
the $\chi^2$
quality of the fit is found by increasing the order of the polynomial
function used.

Once the equation of state function $e(\rho)$ is known, one
can calculate
straightforwardly the surface pressure, defined as
$P(\rho)=\rho^2(\partial e/\partial\rho)$, and the isothermal
compressibility
\begin{equation}
\kappa(\rho)=\frac{1}{\rho}\left(\frac{\partial\rho}{\partial P}\right)
\ .
\label{com}
\end{equation}
{}From the inverse compressibility $1/\kappa$ one obtains the velocity of
sound
\begin{equation}
c(\rho) = \left(\frac{1}{m\rho\kappa}\right)^{1/2} \ .
\label{vs}
\end{equation}
In Fig. 2 we show the comparison between the surface pressure obtained
from our equation of state and the result of Ref. 11. An
appreciable difference is found in the regimes of low and high densities.
A larger difference is found by comparing the predictions of the two fits
for the velocity of sound, shown in Fig. 3. The velocity of sound at the
2D equilibrium density is $c(\rho_0)=(92.8\pm 0.6)$ m/sec,
which is nearly 3 times smaller than the velocity of sound at the
saturation density of 3D bulk liquid $^4$He, $c^{3D}(\rho_0)=238.3$
m/sec.$^{28}$

At $T=0$, and in the limit of a 2D film, the third sound velocity $c_3$
coincides with the velocity of sound $c^{2D}$ of the purely 2D
liquid.$^4$ Then, for low coverages $c^{2D}$ must reproduce quite
accurately the $c_3$ in monolayers adsorbed on strongly binding
substrates. A direct comparison between $c^{2D}$ and experimental $c_3$
for low-coverage films is however difficult by the present
uncertainties in the determination of the liquid coverages.$^{29}$

The density dependence of the
velocity of sound is linear over a wide range of densities.
Only when approaching the spinodal density, where the system becomes
unstable
against density fluctuations, the velocity of sound drops
suddenly to zero
and the compressibility diverges. The estimation of the spinodal
density from our fit gives: $\rho_{sp}=(0.228 \pm 0.002)\sigma^{-2}$,
which is significantly smaller than the value
$\rho_{sp}=0.247\sigma^{-2}$ obtained from the data of Ref. 11. The
spinodal density of 2D homogeneous liquid $^4$He has been estimated also
by Clements {\it et al.}$^{1}$ from the fit of their HNC/EL
variational results
for the equation of state: they find $\rho_{sp}=0.202\sigma^{-2}$.

At the spinodal density $\rho_{sp}=0.228\sigma^{-2}$ the mean
interparticle distance is $r_{sp}=(1/\pi\rho_{sp})^{1/2}=3.02$\AA.
For 3D
bulk liquid $^{4}$He an accurate estimation of the spinodal density
has been given in Ref. 30 with the result $\rho_{sp}^{3D}=(0.264\pm
0.002)\sigma^{-3}$, corresponding to a mean interparticle distance of
$r_{sp}^{3D}=(3/4\pi\rho_{sp}^{3D})^{1/3}=2.47$\AA. It is interesting to
notice that at
the 2D freezing density $\rho_l=0.443\sigma^{-2}$, estimated
in Ref. 11, the mean interparticle
distance is only $r_l=2.17$\AA, whereas at the 2D equilibrium density
the interparticle spacing is $r_{eq}=2.71$\AA. The corresponding values
of the average interatomic distance in the 3D case are respectively:
$r_l^{3D}=2.10$\AA and $r_{eq}^{3D}=2.22$\AA. The mean interparticle
distances of the 2D and
3D liquid $^4$He barely overlap; at freezing the particles are only 3\%
further apart in 2D than in 3D, at the equilibrium density the difference
increases to 22\% and the same difference persists down to the spinodal
density.

Important information on the structure of the ground state is obtained
from the two-body radial distribution function
\begin{equation}
g(r_{12})=\frac{N(N-1)}{\rho^2}\frac{\int|\Phi_0({\bf r}_1,...,
{\bf r}_N)|^2
d{\bf r}_3...d{\bf r}_N}{\int|\Phi_0({\bf r}_1,...,{\bf r}_N)|^2
d{\bf r}_1
...d{\bf r}_N}
\label{gr}
\end{equation}
and from its Fourier transform, the static structure factor
\begin{equation}
S(k) = 1 + \rho\int d{\bf r} e^{i{\bf k}\cdot{\bf r}}(g(r)-1) =
\frac{1}{N} \, \frac{\langle\Phi_0|\rho_{\bf q}\rho_{-{\bf
q}}|\Phi_0\rangle} {\langle\Phi_0|\Phi_0\rangle} \ ,
\label{sk}
\end{equation}
with
\begin{equation}
\rho_{\bf q} = \sum_{i=1}^{N} e^{i{\bf q}\cdot{\bf r}_i} \ .
\label{rq}
\end{equation}
Both these quantities can be calculated using the method for pure
estimators described in Ref. 25.
In Fig. 4 we show the results obtained for the radial distribution
function at three different densities. As the density increases
more peaks
at large interparticle
distances appear. At the highest density shown $\rho=0.420\sigma^{-2}$,
which is just before freezing, four peaks are clearly visible in the
distribution function: a clear indication that the system is close to
solidification. In Table III, we report the position $r_m$ and the
height $g(r_m)$ of the first
peak in the radial distribution function for some of the densities
calculated. The
height of the first peak in $g(r)$ increases with the density and it
shifts towards
smaller interparticle distances. It is interesting to compare the height
of the first peak in the radial distribution function for the 2D and 3D
systems: in 2D at the equilibrium density $g(r_m)\simeq 1.25$ whereas
the corresponding value in 3D is 1.38. This is a clear indication that
the 2D system is more dilute and possesses less correlations at
equilibrium than its 3D counterpart.
Close to the freezing density the heights of the first
peak in $g(r)$ of the 2D and 3D systems become comparable. In fact, as
previously discussed, the mean interparticle distance of the two systems
become similar when freezing is approached.

In Fig. 5 we show the static structure factor for three values of the
density. As the density increases the peak in
$S(k)$ increases and the values at the lowest momenta accessible in our
calculation increase. Due to phonon excitations the form factor $S(k)$
is expected to go to zero in the
long wavelength limit as $S(k)\sim k/2mc$. As the density decreases
and the
spinodal density is approached the velocity of sound $c$ drops to zero
and consequently the slope in $S(k)$ diverges. This behaviour, which has
been observed in the variational calculations of Ref. 1, agrees
qualitatively with our {\it ab initio} calculations.

\subsection{Condensate fraction and momentum distribution}

Another quantity of great interest is the one-body density matrix
$\rho({\bf r}',{\bf r})$, which is related to the change in the
ground-state wavefunction when a particle is removed from position
${\bf r}$ and
replaced at position ${\bf r}'$. For a homogeneous system $\rho(r)$ is
defined as
\begin{equation}
\rho(r) = \langle\Phi_0|\hat{\psi}^{\dagger}(r)\hat{\psi}
(0)|\Phi_0\rangle
=  N \, \frac{\int\Phi_0({\bf r}_1 + {\bf r},...,{\bf
r}_N)\Phi_0({\bf r}_1,...,{\bf r}_N) d{\bf r}_2...d{\bf r}_N}
{\int|\Phi_0({\bf r}_1,...,{\bf r}_N)|^2 d{\bf r}_1...d{\bf
r}_N},
\label{rho}
\end{equation}
where $\hat{\psi}(0)$ and $\hat{\psi}^{\dagger}(r)$ are, respectively,
the
field operators which destroy a particle from position $r=0$ and create
one at position $r$.

In the DMC algorithm the mixed matrix element
$\langle\psi_T|\hat{\psi}^{\dagger}(r)\hat{\psi}(0)|\Phi_0\rangle$,
involving the trial wavefunction $\psi_T$, can be calculated by
averaging over the asymptotic distribution function
$f({\bf R},t\rightarrow
\infty)$ the relative change in the trial wavefunction when a
particle is
displaced from position ${\bf r}_i$ to ${\bf r}_i + {\bf r}$
\begin{equation}
\langle\psi_T|\hat{\psi}^{\dagger}(r)\hat{\psi}(0)|\Phi_0\rangle =
\frac{\int f({\bf R},t\rightarrow\infty)\left( \psi_T({\bf r}_1,...,{\bf
r}_i +
{\bf r},...,{\bf r}_N)/\psi_T({\bf r}_1,...,{\bf r}_N)\right)  d{\bf
r}_1 ...
d{\bf r}_N}{\int f({\bf R},t\rightarrow\infty) d{\bf r}_1 ... d{\bf r}_N}.
\label{mrho}
\end{equation}
The calculation of the ground-state expectation value
$\langle\Phi_0|\hat{\psi}^{\dagger}(r)\hat{\psi}(0)|\Phi_0\rangle$ can not
be obtained straightforwardly from our method for pure estimators because
it involves the knowledge of $\Phi_0({\bf R}')\Phi_0({\bf R})$, where
${\bf R}'$ is the configuration vector with one particle displaced
by $r$, instead of $\Phi_0^2({\bf R})$. To calculate the one-body density
matrix $\rho(r)$ we employ
thus the extrapolation technique described in Sec. II.
For the highest density calculated, $\rho = 0.420 \sigma^{-2}$,
we have used
as importance sampling both the McMillan wavefunction (9) and the one
proposed by Reatto (10). The extrapolation technique gives, within
statistical errors, the same result for $\rho(r)$ in the two cases.

The asymptotic limit of $\rho(r)$ gives the fraction of particles $n_0$
condensed into the zero-momentum state
\begin{equation}
n_0 = \lim_{r\rightarrow\infty}\rho(r) .
\label{nzero}
\end{equation}
In Fig. 6 we show the results for the condensate fraction obtained at
different densities ranging from the spinodal point up to the freezing
density. We have fitted our data with the quadratic polynomial
\begin{equation}
n_0(\rho) = n_0(\rho_0) + a\left(\frac{\rho-\rho_0}{\rho_0}\right)
+ b\left(\frac{\rho-\rho_0}{\rho_0}\right)^2,
\label{nzfit}
\end{equation}
where $\rho_0$ is the equilibrium density $\rho_0=0.284\sigma^{-2}$.
The values of the parameters giving the best fit are the following
\begin{eqnarray}
n_0(\rho_0) &=& 0.233 \pm 0.001 \nonumber \\
a &=& -0.583 \pm 0.006          \\
b &=&  0.44 \pm 0.02 .          \nonumber
\end{eqnarray}
The value for the condensate fraction at equilibrium density
$n_0(\rho_0)$ is consistent with the estimation reported in Ref. 14
$n_0(\rho_0)\simeq 0.22$, obtained by extrapolating to zero temperature
the PIMC results for the algebraic decay of the one-body density matrix.
Our results for the condensate fraction are somewhat smaller than the
ones reported in Ref. 11; for example at low density,
$\rho=0.275\sigma^{-2}$, we find $n_0 = 0.251 \pm 0.005$ whereas the GFMC
calculation of Ref. 11 gives $n_0^{GFMC}=0.36\pm 0.05$. This discrepancy
between DMC and GFMC results persists over the whole density range and
only at very high density, $\rho=0.400\sigma^{-2}$, $n_0^{DMC}$ and
$n_0^{GFMC}$ become consistent. The reason for this is unclear, certainly
it is not due to the revised version of the Aziz potential used in our
DMC simulation. We have repeated our DMC calculation of the condensate
fraction using the Aziz potential for the
two densities $\rho=0.275\sigma^{-2}$ and $\rho=0.400\sigma^{-2}$ and no
difference was found with the results obtained with Aziz II.
It is interesting to notice that near the freezing density, where
the mean
interparticle distance in the 2D and 3D systems are comparable, also the
condensate fraction is nearly the same ($n_0\simeq 4\%$).

By Fourier transforming the one-body density matrix $\rho(r)$, one gets
the momentum distribution of the system
\begin{equation}
n(k) = (2\pi)^2\rho n_0\delta(k) + \rho\int d{\bf r}e^{i{\bf k}\cdot{\bf
r}}(\rho(r)-\rho(\infty)).
\label{nk}
\end{equation}
In Fig. 7 we show the momentum distribution plotted as $kn(k)$ for three
different densities. As the density increases, more and more particles
leave the condensate state and the maximum in $kn(k)$ shifts towards
higher momenta. The shoulder present at high momenta is interpreted in
3D bulk $^4$He as coming from the zero-point motion of the
rotons.$^{31}$ A
similar interpretation could be valid also in the 2D case, in this case
though the shoulder appears more pronounced than in 3D, particularly
at low
densities, and shifted towards smaller $k$'s.

\section{Conclusions}

The properties of homogeneous 2D liquid $^4$He at zero temperature have
been investigated by means of a quadratic diffusion Monte Carlo method.
As interatomic potential we have used a revised version of the Aziz
potential which is expected to be more accurate in describing the
interaction between helium atoms. The energy per particle has been
calculated for a wide range of densities from the spinodal point up to
the freezing density, providing an accurate determination of the equation
of state function. The spinodal density of the 2D system is estimated.
The radial distribution function and the static structure factor are
calculated for various densities employing a recently devised method for
estimating pure expectation values. The fraction of particles in the
condensate state has been calculated for various densities and the
density dependence of the condensate fraction has been also estimated.

\acknowledgements
This work has been supported in part by DGICYT (Spain) Grant No.
PB92-0761
and No. PB90-06131. S. G. acknowledges a post-doctoral grant from the
Ministerio
de Educaci\'on y Ciencia (Spain). The supercomputing facilities provided
by CEPBA and
CESCA are also gratefully acknowledged.

\begin{figure}

\caption{Equation of state for 2D liquid $^4$He. The solid circles
correspond to the DMC energies obtained with the Aziz II potential (the
statistical error bars are smaller than the size of the symbols); the
solid line is the polynomial fit (11) to the calculated energies.
The open diamonds are the GFMC results of Ref. 11 with the Aziz
potential and the dashed line is the reported fit to these values.}

\end{figure}
\begin{figure}

\caption{Density dependence of the surface pressure as obtained from our
fit to the equation of state function (solid line), and from the fit of
Ref. 11 (dashed line).}

\end{figure}
\begin{figure}

\caption{Density dependence of the velocity of sound as obtained from our
fit to the equation of state function (solid line), and from the fit of
Ref. 11 (dashed line).}

\end{figure}
\begin{figure}

\caption{Radial distribution function for three densities:
$\rho=0.275 \sigma^{-2}$ (solid line), $\rho=0.320 \sigma^{-2}$ (short
dashed line), $\rho=0.420 \sigma^{-2}$ (long dashed line).}

\end{figure}
\begin{figure}

\caption{Static structure function for three densities:
$\rho=0.275 \sigma^{-2}$ (solid line), $\rho=0.320 \sigma^{-2}$ (short
dashed line), $\rho=0.420 \sigma^{-2}$ (long dashed line).}

\end{figure}
\begin{figure}

\caption{Density dependence of the condensate fraction. Solid circles
with error bars: results of DMC calculations; solid line: fit from eq.
(20).}

\end{figure}
\begin{figure}

\caption{Momentum distribution for three densities:
$\rho=0.275 \sigma^{-2}$ (solid line), $\rho=0.320 \sigma^{-2}$ (short
dashed line), $\rho=0.420 \sigma^{-2}$ (long dashed line).}

\end{figure}

\begin{table}

\caption{Results for the total and partial energies from DMC calculations.
The potential energies per particle have been obtained from the
calculation of pure expectation values.}

\begin{tabular}{cccc}
$\rho (\sigma^{-2})$  &  $E/N (K)$ & $V/N (K)$ & $T/N (K)$ \\
\tableline
0.235 & -0.8480$\pm$ 0.0016 & -3.850$\pm$ 0.014 & 3.002$\pm$ 0.014  \\
0.255 & -0.8799$\pm$ 0.0020 & -4.234$\pm$ 0.022 & 3.354$\pm$ 0.022  \\
0.275 & -0.8950$\pm$ 0.0019 & -4.680$\pm$ 0.015 & 3.785$\pm$ 0.016  \\
0.320 & -0.8599$\pm$ 0.0017 & -5.722$\pm$ 0.013 & 4.862$\pm$ 0.013  \\
0.340 & -0.7932$\pm$ 0.0031 & -6.213$\pm$ 0.026 & 5.419$\pm$ 0.026  \\
0.380 & -0.5697$\pm$ 0.0047 & -7.390$\pm$ 0.026 & 6.820$\pm$ 0.026  \\
0.420 & -0.1524$\pm$ 0.0047 & -8.532$\pm$ 0.024 & 8.379$\pm$ 0.025  \\
\end{tabular}

\end{table}

\begin{table}

\caption{Parameters of the equation of state (11)
for our DMC results and the GFMC results of Ref. 11}

\begin{tabular}{ccc}
Parameter &  DMC  &  GFMC  \\
\tableline
$\rho_0 (\sigma^{-2})$  &  0.28380 $\pm$ 0.00015  &  0.28458  \\
$e_0  (K)$              & -0.89706 $\pm$ 0.00061  & -0.8357   \\
$B    (K)$              &  2.065   $\pm$ 0.014    &  1.659    \\
$C    (K)$              &  2.430   $\pm$ 0.035    &  3.493    \\
$\chi^2/\nu$            &  0.99                   &  1.45     \\
\end{tabular}

\end{table}

\begin{table}

\caption{Position $r_m$ and height $g(r_m)$ of the first peak in the
radial distribution function}

\begin{tabular}{ccc}
$\rho (\sigma^{-2})$  &  $r_m$ (\AA)  &  $g(r_m)$  \\
\tableline
0.235    &   4.09   &   1.209 $\pm$ 0.005  \\
0.255    &   4.03   &   1.233 $\pm$ 0.005  \\
0.275    &   3.98   &   1.253 $\pm$ 0.002  \\
0.320    &   3.87   &   1.323 $\pm$ 0.003  \\
0.340    &   3.84   &   1.363 $\pm$ 0.009  \\
0.380    &   3.63   &   1.426 $\pm$ 0.013  \\
0.420    &   3.61   &   1.522 $\pm$ 0.006  \\
\end{tabular}

\end{table}

\end{document}